\documentclass[conference]{IEEEtran}
\IEEEoverridecommandlockouts
\usepackage{cite}
\usepackage{amsmath,amssymb,amsfonts}
\usepackage{algorithmic}
\usepackage{graphicx}
\usepackage{textcomp}
\usepackage{xcolor}
\usepackage{hyperref} 
\def\BibTeX{{\rm B\kern-.05em{\sc i\kern-.025em b}\kern-.08em
    T\kern-.1667em\lower.7ex\hbox{E}\kern-.125emX}}
\begin{document}

\title{Qlippy: A Retrieval-Augmented GenAI Assistant for Reproducible Quantum Workflows and Experiment Tracking\\
\thanks{This work has been supported by Business Finland projects EM4QS (155/31/2024) and the Research Council of Finland project DEQSE (349945).}
}

\author{\IEEEauthorblockN{Mahee Gamage}
\IEEEauthorblockA{
\textit{University of Jyväskylä}\\
Jyväskylä, Finland \\
mahee.s.hewagamage@jyu.fi}
\and
\IEEEauthorblockN{Vlad Stirbu}
\IEEEauthorblockA{
\textit{University of Jyväskylä}\\
Jyväskylä, Finland \\
vlad.a.stirbu@jyu.fi}
}

\maketitle

\begin{abstract}
Quantum software development is iterative and error-prone. Noisy hardware and repeated re-execution make experiment tracking, provenance, and reproducibility essential, yet these practices are hard to adopt because of tooling complexity and the specialized knowledge they demand. General-purpose language models can help but tend to hallucinate and lack grounding in domain-specific tooling. We present Qlippy, a retrieval-augmented GenAI assistant embedded in the development environment that grounds its responses in a curated corpus of quantum-software-engineering knowledge. Qlippy explains reproducibility and provenance concepts in context and augments existing Qiskit programs with MLflow-based experiment tracking aligned to the QProv schema. By separating knowledge from model parameters, grounding gives explicit control over the scope and provenance of the assistant's responses and reduces reliance on model scale, which points toward low-cost, privacy-preserving local deployment.
\end{abstract}


\section{Introduction}

Quantum computing is emerging as an interdisciplinary field that attracts practitioners from physics, mathematics, and computer science across research, industry, and education. As access to quantum programming frameworks and cloud-based quantum hardware expands, more developers are building and running quantum algorithms in practice. However, the development of quantum software is not only a matter of algorithm design, but also involves practices rooted in modern software engineering, such as experiment tracking, reproducibility, and structured data management. These practices are essential for ensuring that experimental results can be validated, compared, and iteratively improved.%

Despite the wide adoption of these practices in classical computing, adopting them in quantum software development remains challenging. Practitioners entering the field often possess strong theoretical backgrounds but have limited exposure to software engineering concepts, particularly those related to reproducibility and lifecycle management, an issue that is especially pronounced for novice developers. As a result, experiment tracking and provenance collection are either applied superficially or omitted entirely, leading to ad hoc development practices and reduced transparency in experimental results. Furthermore, the complexity of integrating external tools into development workflows increases cognitive load and creates friction for developers. While recent advances in generative AI offer opportunities to support this work, generic language models lack grounding in domain-specific materials and may produce explanations or code that are misaligned with domain practices, limiting their effectiveness in specialized settings.

In this paper, we present a retrieval-augmented GenAI assistant integrated into Visual Studio Code, designed to support the adoption of reproducible quantum software practices. The assistant leverages curated knowledge sources, including tools documentation and relevant research literature, to provide context-aware explanations and code augmentation that help developers apply experiment tracking and collect structured provenance information. Through a guided scenario, we demonstrate how retrieval-augmented generation can bridge the gap between quantum algorithm development and modern software engineering practices, reducing cognitive overhead while reinforcing understanding through immediate application.

\section{Background}

Experiment tracking is particularly important in quantum software development due to the intrinsic instability of current hardware, where noise, calibration drift, and backend variability can significantly affect results across executions. These conditions make quantum development inherently iterative, requiring practitioners to repeatedly refine circuits, parameters, and execution strategies while carefully observing marginal improvements \cite{kinanen2025toolchain}. In this context, classical machine learning tools such as MLflow \cite{zaharia2018accelerating} provide a structured way to log parameters, results, and artifacts, supporting reproducibility and comparative analysis. However, despite its importance, experiment tracking is rarely adopted systematically in quantum software development, leaving practitioners to rely on ad hoc practices.

Prior work on laboratory practices and reporting related to quantum software experiments \cite{moguel2025quantum}, highlights comprehensive approaches to structuring experiments and reporting results, yet also illustrates that the current bar for systematic tracking and documentation remains high for novice developers. This gap reinforces the need for more accessible, integrated approaches that embed experiment tracking into everyday development workflows rather than treating it as an advanced or optional practice. Beyond basic logging, effective experiment tracking in quantum software requires structured metadata and detailed execution context, including circuits, compilation steps, backend settings, and noise conditions. Without this, results are difficult to reproduce or compare across runs. The QProv schema addresses this by proposing a comprehensive attribute schema on how provenance is captured, enabling more reliable and interpretable experimentation \cite{weder2021qprov}.

Adopting these practices is challenging in its own right: developers must reason simultaneously about quantum algorithms, hybrid execution environments, and supporting infrastructure, while tools such as MLflow and provenance approaches like QProv add further layers of abstraction \cite{reinikainen2024qcforall}. These are experiential skills that are difficult to acquire from documentation alone, which motivates tool-supported approaches embedded directly in the development workflow.

Generative AI, particularly Large Language Models (LLMs), is increasingly used as an assistant in software engineering tasks due to its ability to interpret natural language and generate contextually relevant responses \cite{brown_language_2020}. However, LLMs rely on parametric memory encoded during pre-training, which leads to well-documented limitations such as hallucinations, outdated knowledge, and lack of grounding in domain-specific facts \cite{lewis_retrieval-augmented_2020}. These issues are particularly problematic in specialized domains like quantum software development, where correctness and traceability are essential. Retrieval-Augmented Generation (RAG) mitigates these limitations by integrating external, curated knowledge sources into the generation process, allowing the model to retrieve relevant information and ground its responses in verifiable content \cite{gao_retrieval-augmented_2024}. By separating knowledge from model parameters and enabling controlled, domain-specific knowledge bases, RAG supports more reliable, transparent, and domain-appropriate interactions.

\section{Design}

\subsection{System Architecture}

\begin{figure}
    \centering
    \includegraphics[width=\linewidth]{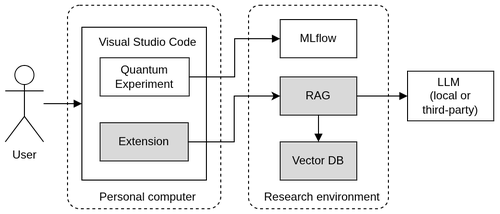}
    \caption{The system architecture (newly developed components in gray)}
    \label{fig:system-architecture}
\end{figure}

The system architecture comprises two distinct environments interconnected via a REST API, as illustrated in Fig.~\ref{fig:system-architecture}. On the developer's local workstation, a Visual Studio Code extension provides a chat-based interface, enabling natural language interaction with the assistant.

The research environment hosts two primary services: an MLflow tracking server for managing experimental metadata and a RAG (Retrieval-Augmented Generation) service. Implemented in Python using FastAPI\footnote{\href{https://fastapi.tiangolo.com/}{https://fastapi.tiangolo.com/}} and orchestrated via LangChain\footnote{\href{https://www.langchain.com/}{https://www.langchain.com/}}, the RAG service operates through three core components. First, a ChromaDB\footnote{\href{https://www.trychroma.com/}{https://www.trychroma.com/}} vector database stores curated documentation on quantum software experiment tracking, serving as the system's non-parametric memory. The corpus comprises five research papers, two synthetically authored reference documents, and a curated subset of the MLflow documentation, ingested via a pipeline consisting of file parsing, chunking, and embedding stages. Second, a retriever identifies and extracts relevant context from this database based on user input. Finally, a generator synthesizes the retrieved context and the user query to produce natural-language explanations or modified code segments. The service is designed to be LLM-agnostic, ensuring seamless interoperability with both commercial APIs and local open-weight models without requiring modifications to the core pipeline.

\subsection{Developer interaction model}

The system follows an interaction loop tightly integrated into the developer workflow. A developer begins by writing or exploring quantum code within the development environment and then poses a question to the assistant in natural language. In response, the assistant performs three complementary actions: it explains the underlying concept, augments or modifies the code where appropriate, and references relevant sources to ground the explanation. Because each explanation is coupled with an immediate code-level change delivered in context, conceptual understanding and practical implementation reinforce each other rather than being deferred to external documentation or tutorials.

\subsection{Integration with Tracking}

The assistant integrates directly with MLflow as a real-world experiment tracking system, ensuring that the generated outputs align with tools and practices expected in professional and research settings. Instead of introducing a simplified or simulated mechanism, the system augments user code by automatically inserting MLflow-based tracking calls, enabling the recording of user's experiment runs, parameters, metrics, and artifacts in the same manner as used in research and industry environments. This makes the resulting workflow directly transferable to real-world practice.

\section{Demonstration}

\subsection{Advisory Scenario: In-Context Explanation of Experiment Tracking}

\begin{figure}
    \centering    \includegraphics[width=\linewidth]{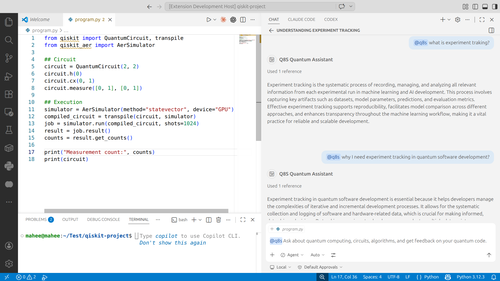}
    \caption{Qlippy in the advisory scenario for experiment tracking.}
    \label{fig:education}
\end{figure}

The first demonstration scenario focuses on advisory support, where the artifact helps quantum software practitioners, including novices, acquire knowledge about experiment tracking concepts without requiring prior exposure to tools such as MLflow or provenance models like QProv. In this scenario, the user interacts with the system through natural language queries embedded within the development environment (e.g., VSCode), depicted in Fig.~\ref{fig:education}. The system retrieves relevant information from curated sources, including MLflow documentation and QProv specifications, and generates context-aware responses that explain both conceptual and practical aspects of experiment tracking.

This interaction is iterative and exploratory: users can refine questions, request examples, and connect abstract provenance concepts with concrete implementation practices. The system effectively acts as a domain-specific advisor, bridging the gap between quantum software development expertise and experiment tracking practices. The success of this scenario is demonstrated when the system provides coherent, grounded explanations that enable users to understand what to track (e.g., circuit properties, execution parameters) and why these are important in the NISQ context. This directly addresses the identified knowledge barrier and supports understanding through interaction rather than static documentation.

\subsection{Augmentation Scenario: AI-Assisted Injection of Experiment Tracking Code}

\begin{figure}
    \centering
    \includegraphics[width=\linewidth]{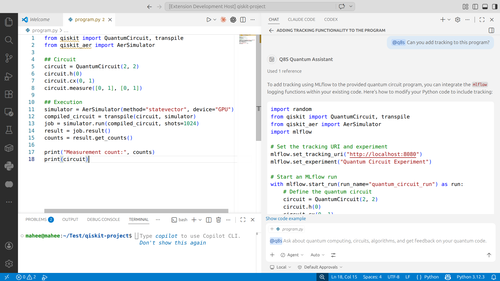}
    \caption{Using the Qlippy in the injection of experiment tracking code}
    \label{fig:augmentation}
\end{figure}

The second demonstration scenario focuses on augmentation, where the artifact actively modifies existing quantum programs to include experiment tracking capabilities. In this case, the user provides a functional Qiskit-based quantum program along with a natural language instruction describing the tracking requirements (e.g., aligning with QProv attributes). The system retrieves the necessary knowledge and generates a modified version of the program with integrated MLflow logging.

Unlike the advisory scenario, which supports understanding, this scenario reduces the need for manual implementation by automating the translation of conceptual knowledge into executable code. The generated output preserves the original program logic while augmenting it with tracking functionality, enabling the recording of parameters, metrics, and provenance data during execution. The demonstration is considered successful when the modified program executes correctly and logs experiment data consistent with QProv guidelines into the MLflow tracking system.

This scenario illustrates how the artifact shifts from a passive advisory role to an active development assistant, lowering the operational barrier to adopting structured experiment tracking in quantum workflows.

\section{Discussion}

\subsection{Preliminary evaluation}

We conducted a preliminary evaluation of the advisory scenario using the RAGAS framework \cite{es2024ragas} (an LLM-as-a-Judge--based
evaluation approach), over 20 question--answer pairs spanning five subdomains of experiment tracking and four question types (factual, summary, reasoning, and unanswerable), with \texttt{GPT-5.4-mini} as judge\footnote{OpenAI API, accessed April 2026.}. Responses were scored on faithfulness, context recall, answer relevancy, and factual correctness (recall value based), excluding unanswerable questions from the means since evaluation metrics are not designed to meaningfully evaluate that question type. Across four generator models (\texttt{GPT-4o-mini} and \texttt{GPT-5.4} via the OpenAI API; \texttt{Gemma4:e4b} and \texttt{Qwen3.5:9b}, quantized and served locally via Ollama), the locally deployable \texttt{Qwen3.5:9b} achieved the highest answer relevancy (0.917), while \texttt{GPT-5.4} achieved the highest factual correctness (0.908), only 12\% higher than \texttt{Qwen3.5:9b}'s (0.814). Given the scale of the question set and the use of an LLM as judge, these results are best read as relative comparisons between configurations. A more rigorous evaluation is needed before drawing firm conclusions about system quality.

\subsection{Benefits for quantum software development}

A retrieval-augmented assistant embedded in the development environment offers several practical benefits for adopting reproducible quantum software practices. First, it lowers the expertise barrier: developers can apply experiment tracking and provenance collection without deep familiarity with MLflow, the QProv schema, or the surrounding DevOps tooling. Second, support is delivered in context and just in time, at the point in the workflow where a practice becomes relevant, which reduces the friction of adopting unfamiliar tools. Third, and in contrast to general-purpose coding assistants, grounding the model in a curated corpus gives explicit control over the scope and provenance of its knowledge, keeping responses aligned with a target practice rather than the open-ended behavior of an ungrounded model. Education is one setting where these benefits apply directly, helping instructors introduce advanced practices without mastering the full tooling ecosystem, but the same properties support research and industry teams onboarding reproducible workflows.

\subsection{Operational concerns}

A central consideration in deploying such an assistant is the tradeoff between cost, performance, and control over the execution environment. As the preliminary evaluation indicates, locally hosted models such as \texttt{qwen3.5:9b} can approach commercial offerings like \texttt{GPT-5.4} on answer relevancy and factual coverage when used within a Retrieval-Augmented Generation (RAG) pipeline. This is largely because RAG shifts the burden of knowledge from the model's parametric memory to curated external sources, effectively reducing the role of the generator to synthesis and contextualisation. As a result, the advantage of large parametric models is less pronounced in such constrained, domain-specific scenarios.

From a cost perspective, local deployment eliminates recurring API usage fees, which can become significant under continuous or large-scale usage. It also provides greater control over data flows, addressing privacy concerns that are especially salient when proprietary code or unpublished research artifacts are involved. These properties make locally hosted solutions attractive across research, industry, and teaching settings alike, wherever cost, confidentiality, or control over the execution environment are priorities.

However, these benefits come with their own tradeoffs, including the need to provision and maintain adequate computational infrastructure, manage model updates, and ensure system reliability. The decision between commercial and local models is therefore not only a performance comparison but a broader socio-technical choice involving cost structures, governance requirements, and operational capabilities.

\subsection{Threats to validity}

This study has several limitations, identified according to Wohlin et al \cite{wohlin2012experimentation} as follows. First, the evaluation is based on a prototype implementation, which demonstrates feasibility but may not reflect performance and usability in real-world deployments. Second, the system depends on curated knowledge sources; any gaps or biases in these sources directly affect retrieval quality and response correctness. Third, the artifact has not been extensively validated with real users and development teams, limiting conclusions about its adoption in practice. Finally, despite grounding through RAG, residual LLM limitations remain, including occasional hallucinations and variability in responses, particularly for complex queries.

\section{Conclusion}

This paper presented Qlippy, a retrieval-augmented GenAI assistant that embeds experiment tracking and provenance practices directly into the quantum software development workflow. By grounding its responses in curated sources and coupling explanations with automated MLflow instrumentation aligned to the QProv schema, the assistant helps developers apply reproducible practices as they write code, one use case being novice practitioners learning these concepts for the first time. Our preliminary evaluation indicates that, within a RAG pipeline, a locally hosted model can approach a commercial one on answer relevancy and factual coverage, making low-cost, privacy-preserving deployment realistic across research, industry, and teaching settings alike. Since the evaluation remains preliminary, future work will extend the question set with multi-expert validation, add a formal assessment of provenance completeness for the augmentation scenario, and compare the grounded assistant against an ungrounded model of equal size and against general-purpose coding assistants.

\bibliographystyle{IEEEtran}
\bibliography{bibliography}

\end{document}